\documentclass{article}
\usepackage{ijcai26}
\makeatletter
\renewcommand\section{\@startsection{section}{1}{\z@}%
  {-1.0ex plus -0.2ex minus -0.2ex}%
  {0.8ex plus 0.2ex}%
  {\normalfont\large\bfseries}}
\renewcommand\subsection{\@startsection{subsection}{2}{\z@}%
  {-0.8ex plus -0.2ex minus -0.2ex}%
  {0.6ex plus 0.2ex}%
  {\normalfont\normalsize\bfseries}}
\renewcommand\subsubsection{\@startsection{subsubsection}{3}{\z@}%
  {-0.6ex plus -0.2ex minus -0.2ex}%
  {0.4ex plus 0.2ex}%
  {\normalfont\normalsize\bfseries}}
\makeatother

\usepackage{soul}
\usepackage{url}
\usepackage[hidelinks]{hyperref}
\usepackage[utf8]{inputenc}
\usepackage[small]{caption}
\usepackage{graphicx}
\usepackage{amsmath}
\usepackage{amsthm}
\usepackage{booktabs}
\usepackage{multirow}
\usepackage{algorithm}
\usepackage{algorithmic}
\usepackage[switch]{lineno}
\usepackage{enumitem}
\usepackage{titlesec}
\usepackage{amssymb}

\nolinenumbers

\titleformat{\section}
  {\normalfont\bfseries\large}
  {\thesection}
  {1em}
  {}

\titleformat{\subsection}
  {\normalfont\bfseries}
  {\thesubsection}
  {1em}
  {}

\titleformat{\subsubsection}
  {\normalfont\bfseries}
  {\thesubsubsection}
  {1em}
  {}

\title{Next Point-of-interest (POI) Recommendation Model Based on Multi-modal Spatio-temporal Context Feature Embedding }

\author{
Lingyu Zhang$^1$
\and
Pengfei Xu$^2$\and
Rui Ban$^{3}$\and
Zhenchao Zhang$^{3}$\and
Songtao Liu$^{3}$\and
Yan Wang$^{2}$\And
Yunhai Wang$^4$\\
\affiliations
$^1$Institute of Trustworthy Autonomous Systems, Southern University of Science and Technology(SUSTech), Shenzhen, China.\\
$^2$School of Information Sciences and
Technology, Northwest University, Xi’an, China\\
$^3$China Information Technology Designing \& Consulting Institute Co., Ltd., Beijing, 100048, China\\
$^4$Renmin University of China Beijing, China\\
\emails
zhangly6@mail.sustech.edu.cn, 
pfxu@nwu.edu.cn,
\{banrui1, zhangzc131, liust313\}@chinaunicom.cn,
202433687@stumail.nwu.edu.cn, 
wang.yh@ruc.edu.cn
}

\begin{document}

\maketitle

\begin{abstract}
Predicting the next pickup location of individual users is a fundamental problem in intelligent mobility systems, which requires modeling personalized travel behaviors under complex spatiotemporal contexts. Existing methods mainly learn sequential dependencies from raw trajectories, but often fail to capture high-level behavioral semantics and to effectively disentangle long-term habitual preferences from short-term contextual intentions.
In this paper, we propose a semantic-embedding-based dual-stream spatiotemporal attention model for next pickup location prediction. Raw trajectories are first transformed into semantically enriched activity sequences to encode users’ stay behaviors and movement semantics. A dual-stream architecture is then designed to explicitly decouple long-term historical patterns and short-term dynamic intentions, where each stream employs spatiotemporal attention mechanisms to model dependencies at different temporal scales. To integrate heterogeneous contextual information, a context-aware dynamic fusion module adaptively balances the contributions of the two streams. Finally, an attention-based matching strategy is used to predict the probability distribution over candidate pickup locations.
Experiments on real-world ride-hailing datasets demonstrate that the proposed model consistently outperforms state-of-the-art methods, validating the effectiveness of semantic trajectory abstraction and dual-stream spatiotemporal attention for individualized mobility behavior modeling.
\end{abstract}

\section{Introduction}

With the acceleration of urbanization and the rapid development of information technology, intelligent mobility services represented by ride-hailing have become deeply integrated into modern urban life. These services not only greatly enhance the convenience of residents’ travel, but also provide a solid data foundation for building more efficient and smarter urban transportation systems. In this context, accurate travel demand prediction has become a core technological engine for improving platform operational efficiency, optimizing resource allocation, and enhancing user experience\cite{zhang2019multitask}. This problem is typically studied at both the macroscopic regional level and the microscopic individual level, which complement each other and jointly support the operation of intelligent mobility services. Macroscopic prediction focuses on the overall distribution and evolution of urban traffic flows, providing decision support for capacity planning and regional dispatching; microscopic prediction, in contrast, delves into the smallest unit of service—the individual passenger.
This paper focuses on microscopic-level passenger travel demand prediction, specifically predicting the pickup location of a particular user’s next trip. This problem presents challenges that are fundamentally different from those of macroscopic prediction, mainly because an individual passenger’s choice of pickup location is a dynamic decision-making process driven by the coupling of complex internal and external factors. Internal factors such as users’ commuting patterns and lifestyle habits intertwine with external environmental factors such as weather, traffic conditions, and holidays, resulting in highly personalized behavior with strong contextual dependence. Such complex patterns make it a core scientific challenge to accurately predict a user’s next specific departure location from relatively sparse historical travel records. To address these challenges, extensive explorations have been conducted in both academia and industry.
However, a careful examination of existing studies reveals several key limitations in meeting diverse business requirements and in finely modeling individual behavior. Current research mainly focuses on computation-intensive deep learning architectures that improve prediction accuracy by capturing high-order spatiotemporal dependencies, while systematic studies on computationally efficient lightweight models suitable for low-latency deployment are lacking. The few existing lightweight attempts, such as approaches based on Bayesian inference \cite{fei2011bayesian}, gradient boosting decision trees\cite{sun2020gbdt}, or support vector machines\cite{xu2017svm}, typically construct unified, non-personalized models. Due to the lack of effective mechanisms for integrating heterogeneous information beyond users’ historical trajectories, these methods inherently fall short in achieving high-precision personalized prediction. Secondly, mainstream deep learning models still require further refinement in fine-grained representations of individual behavior. Although long short-term memory networks (LSTM) \cite{zhu2017timelstm,huang2019stlstm}, Transformers \cite{yang2022getnext,luo2021stan}, and graph neural networks (GNN) \cite{luo2021stan,naqvi2017influxdb} have been widely applied, they still face challenges in two aspects: first, the feature modeling of spatiotemporal context is not sufficiently fine-grained, making it difficult to capture detailed spatiotemporal interaction patterns; second, the mechanisms for fusing heterogeneous internal and external factors that drive individual decision-making remain incomplete.
These shortcomings make it difficult for models to accurately characterize users’ complex decision logic under specific contexts, thereby constraining further improvements in prediction performance. In view of the above research gaps, this chapter aims to construct a behavior-pattern-driven, multi-model prediction framework that integrates internal and external factors, with the goal of jointly addressing the differentiated business requirements of ride-hailing platforms for high timeliness and high-accuracy prediction. This paper proposes a dual-stream spatiotemporal attention model based on semantic embeddings, aiming to overcome the inherent limitations of existing deep learning methods in coarse-grained spatiotemporal feature modeling and insufficient heterogeneous information fusion. The model architecture begins with the semantic abstraction of raw trajectory data: first, DBSCAN clustering is employed to adaptively discretize continuous geographic space, which effectively removes noisy signal points while transforming raw GPS coordinate sequences into meaningful location cluster sequences. Subsequently, temporally continuous check-in points belonging to the same cluster are aggregated into a single “spatiotemporal event,” and their activity durations are computed, thereby elevating the raw trajectory from a simple point sequence to an activity sequence with rich semantic hierarchy.
On this basis, the model decouples long-term historical trajectories that encode users’ stable patterns from short-term trajectories that reflect recent contextual influences. For these two types of subsequences, the model separately constructs user trajectory embeddings and spatiotemporal relation embeddings, and deeply aggregates their internal spatiotemporal dependencies through self-attention mechanisms. Its core innovation lies in a context-aware dynamic fusion module: this module introduces external knowledge such as functional location attributes and traffic conditions to dynamically compute the fusion weights of long- and short-term feature representations—for example, assigning higher weight to long-term patterns in commuting scenarios, while relying more on short-term context when exploring new areas. Finally, the fused comprehensive feature representation is fed into an attention-based matching network to compute the probability distribution over all candidate locations of being the next travel destination. In this way, the model achieves a fine-grained characterization of individual travel decision-making processes, significantly improving prediction accuracy and model interpretability, and providing solid and actionable decision support for complex tasks such as high-precision capacity dispatching and dynamic pricing on ride-hailing platforms.
The contributions of this paper to the design and optimization of pickup location prediction models are mainly reflected in the following aspect: it designs a dual-stream spatiotemporal attention model based on semantic embeddings. Through trajectory semantic abstraction, decoupling of long- and short-term dependencies, and a context-aware dynamic fusion mechanism, the model achieves fine-grained modeling of individual travel decision-making processes and significantly improves prediction performance in high-accuracy scenarios.

\section{Review of Domestic and International Research}

Before the rise of deep learning, POI prediction mainly relied on traditional statistical models. Among them, Markov models were widely applied. For example, Ashbrook and Starner \cite{ashbrook2002gps} as well as Gambs et al.\cite{gambs2012markov}identified key locations from GNSS trajectories and constructed user-specific Markov models for POI prediction. Such models treat locations as states and perform prediction by computing transition probabilities between states. However, they can only capture short-range and simple spatiotemporal dependencies, making it difficult to effectively model complex user mobility patterns. To address this limitation, Rendle et al. \cite{rendle2010fpmc}
 proposed the Factorizing Personalized Markov Chains (FPMC) model, which incorporates factors such as individual and collective preferences when computing transition probabilities. This approach improves prediction performance to some extent, but it still suffers from limitations in handling long-range spatiotemporal dependencies.
With the continuous development of deep learning techniques, their strong capabilities in feature learning and complex pattern modeling have quickly made them dominant in this field. Recurrent neural networks (RNNs) and their variant, long short-term memory networks (LSTMs), became the mainstream research direction. Many early next-POI recommendation models were built upon RNNs. For instance, STRNN \cite{liu2016strnn}explicitly incorporated the time interval and spatial distance between two consecutive visits into an RNN, achieving certain performance improvements in scenarios such as public safety assessment. SERM \cite{yao2017serm}
 innovatively jointly learned temporal context and semantic context (e.g., POI categories), and modeled users’ temporal regularity and functional preferences through a two-layer RNN architecture. DeepMove \cite{feng2018deepmove} combined an attention layer for learning long-term periodicity with a recurrent layer for learning short-term sequential patterns, using the attention layer to capture long-term periodicity and the RNN to learn short-term sequence patterns, and learning from highly correlated trajectories.
However, these RNN-based models generally suffer from insufficient ability to capture potential associations between non-adjacent locations, and their spatiotemporal modeling is relatively coarse. To overcome the long-distance dependency limitations of RNNs, researchers introduced attention mechanisms into the models. Time-LSTM \cite{zhu2017timelstm} introduced time-gated units into LSTM, dynamically adjusting memory cell weights according to visit timestamps, thereby reducing nighttime prediction errors by 18\%. ATST-LSTM \cite{huang2019stlstm} was the first to introduce a self-attention mechanism into LSTM, generating dynamic weights for each check-in. Visualization results showed that it assigned more reasonable weights to high-frequency paths, but it was still limited to local attention within continuous sequences.
Inspired by the Transformer architecture in natural language processing, self-attention models have demonstrated stronger capabilities in capturing global associations in trajectory modeling. GeoSAN \cite{lian2020geographyaware} was the first to apply self-attention to location recommendation, allowing any two check-in points in a trajectory to interact directly. However, this model adopted fixed-grid spatial discretization and ignored the explicit modeling of temporal and spatial intervals, making it difficult to accurately capture precise distances. Subsequently, STAN \cite{luo2021stan} addressed the shortcomings of GeoSAN by proposing a dual spatiotemporal attention architecture. The first attention layer aggregates spatiotemporal features of all check-ins within a trajectory, while the second layer incorporates personalized item frequency (PIF) to model repeated visit preferences, achieving Recall@k improvements of 10\%–17\% over baselines on four real-world datasets.
In recent years, research has gradually shifted from single spatiotemporal features to the fusion of multimodal contextual information. Early studies captured users’ short-term and stable preferences by defining long-term and short-term check-in trajectories, where LSTM was often used together with long–short-term trajectory modeling \cite{sun2020lstpm,wang2022stneighbourhood,yang2022getnext}
. In addition, to improve recommendation effectiveness, many studies focused on incorporating environmental information such as spatiotemporal context into POI recommendation, demonstrating that factors such as category, check-in time, and geographic location are effective for POI recommendation. These factors were jointly fed into a single LSTM model, yielding significant results. However, as the input sequence length increases, LSTM has limited ability to associate long-distance information. Therefore, recent studies have combined attention mechanisms with LSTM models to balance the weights assigned to recent and distant information, enabling models to capture relevant temporal dependencies across the entire sequence \cite{feng2018deepmove,li2020hierarchical}. The success of attention mechanisms further inspired the introduction of multi-head self-attention (MHSA), which uses multiple layers of self-attention to learn multiple sets of relationships within the input sequence \cite{vaswani2017attention}. MHSA-based models have achieved state-of-the-art performance in POI recommendation tasks \cite{xue2021mobtcast,sun2022tcsanet}, but their potential for location prediction under continuously recorded mobility trajectories remains to be further explored. Other studies have adopted context-aware MHSA neural networks \cite{hong2023context}, integrating information such as multi-scale land-use features of POIs, dwell time, and visit time, achieving an Acc@1 of 45.6\% on GNSS trajectory datasets, which verifies the significant performance gains brought by multimodal contextual information.
In addition, to address issues such as sparse data cold start, limited interpretability, and adaptation to dynamic environments, researchers abroad have actively explored frontier directions such as meta-learning and few-shot learning, graph neural networks (GNNs) and knowledge graphs, as well as online learning and dynamic adaptation. For example, in the area of meta-learning and few-shot learning, meta-learning-based approaches are used to quickly initialize user preferences to address recommendation challenges for new users or low-frequency POIs. In the direction of graph neural networks and knowledge graphs, structured knowledge such as POI semantics and transportation networks is integrated into models to enhance interpretability. In the field of online learning and dynamic adaptation, real-time data-stream-driven model update mechanisms are designed to capture the temporal evolution of urban functions.
Although existing models have made significant progress in capturing spatiotemporal features and multimodal information, several challenges remain to be addressed. First, existing methods often exhibit low flexibility and accuracy when handling complex user trajectories with long time intervals and cross-scenario transitions, especially when modeling personalized user preferences, where the deep fusion of various environmental and contextual information has not been fully exploited. Second, the application of multi-head self-attention models in dynamic prediction tasks is still in the exploratory stage, particularly lacking task-specific optimization for continuous trajectory prediction and location recommendation. Therefore, this study will focus on improving the accuracy of POI recommendation through multimodal information fusion and cross-domain modeling, and on enhancing the ability to capture long-range dependency information through innovative self-attention mechanisms, thereby filling the gaps of existing methods in continuous trajectory prediction.

\section{Semantic-Embedding-Based Dual-Stream Spatio-Temporal Attention Model}

\begin{figure*}[t]
  \centering
  \includegraphics[width=0.95\textwidth]{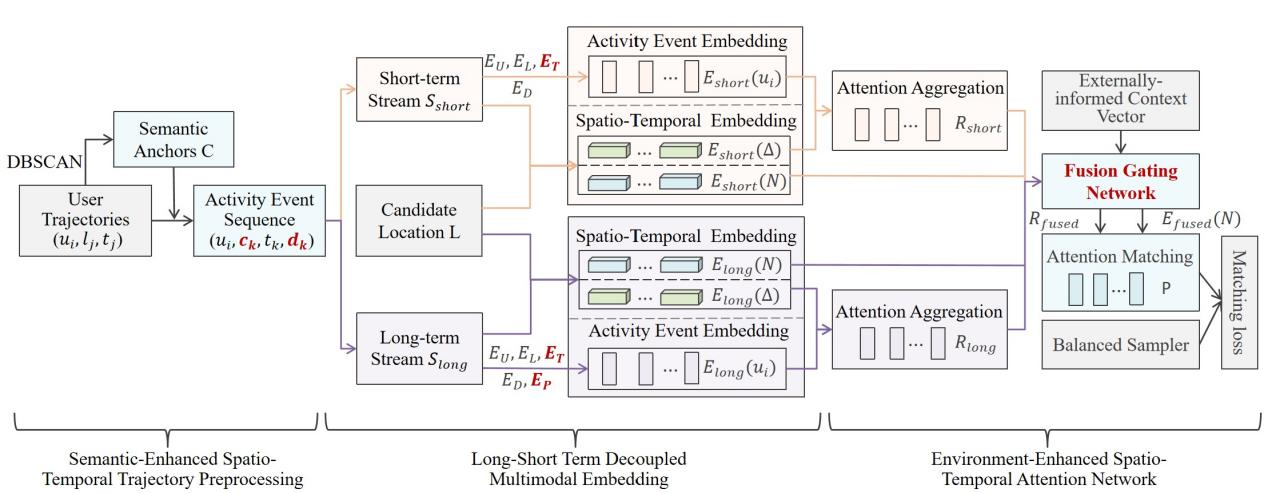}
  \caption{Model architecture of the proposed method.}
  \label{fig:model}
\end{figure*}

This paper proposes a dual-stream spatiotemporal attention model based on semantic embeddings. The core design philosophy of the model is to finely characterize users’ travel decision-making processes from three aspects—semantics, multi-scale modeling, and environmental feature fusion—in order to achieve a significant improvement in prediction performance. Specifically, the overall framework of the model consists of the following components in sequence: first, semantic-enhanced spatiotemporal trajectory preprocessing is performed to transform low-level GPS sequences into activity event sequences enriched with spatiotemporal semantic information; second, long- and short-term decoupled multimodal embeddings are adopted to separately capture users’ stable travel habits and recent immediate intentions, thereby avoiding pattern confusion; finally, an environment-feature-enhanced spatiotemporal attention network is used to dynamically fuse internal trajectory dependencies with external environmental features (such as weather and traffic conditions). The architecture of the model is illustrated in Figure~\ref{fig:model}
\subsection{Semantically Enhanced Spatio-Temporal Trajectory Preprocessing}
This section describes the first stage of the model framework: transforming raw GPS trajectories into structured and semantically rich activity event sequences. As the input stage of the deep learning model, this preprocessing step is a prerequisite for fine-grained modeling. Raw GPS sequences not only contain noise and redundancy, but also lack direct representations of high-level behaviors such as user “stays” and “activities.” Without proper processing, this would inevitably reduce learning efficiency and weaken prediction accuracy. To address this issue, this section introduces a preprocessing pipeline that combines spatial discretization and spatiotemporal event aggregation, refining low-level sensor readings into high-level activity events. This provides high-quality, information-dense inputs for the subsequent spatiotemporal attention network. The specific steps are as follows:
\subsubsection{}
 \begin{enumerate}[label=(\arabic*)]
\item \textbf{Spatial discretization and semantic anchor discovery.}

The goal of this step is to transform continuous geographic space into a set of discrete functional regions, referred to as semantic anchors. Specifically, the set of all users’ historical check-in locations $L = \{l_1, l_2, \ldots, l_L\}$
 is taken as a whole input, and the DBSCAN algorithm is employed to identify latent hotspot regions that embody collective behavioral patterns, thereby forming a discrete cluster set , $C = \{c_1, c_2, \ldots, c_K\}$that represents the distribution of the original geographic coordinates. DBSCAN is chosen for its notable theoretical advantages. First, it can discover clusters of arbitrary shapes distributed along streets or buildings, which closely aligns with the natural forms of urban functional areas. Second, it does not rely on a predefined number of clusters, but instead adaptively performs clustering based on the density distribution of the data. Third, it incorporates a noise identification mechanism that can eliminate outliers caused by signal drift or sporadic visits. This operation not only successfully achieves adaptive discretization of continuous space and significantly optimizes system performance, but more importantly, it aggregates massive coordinate points into a finite number of representative functional regions, accomplishing an initial semantic abstraction at the spatial level.
 \begin{figure*}[t]
  \centering
  \includegraphics[width=0.95\textwidth]{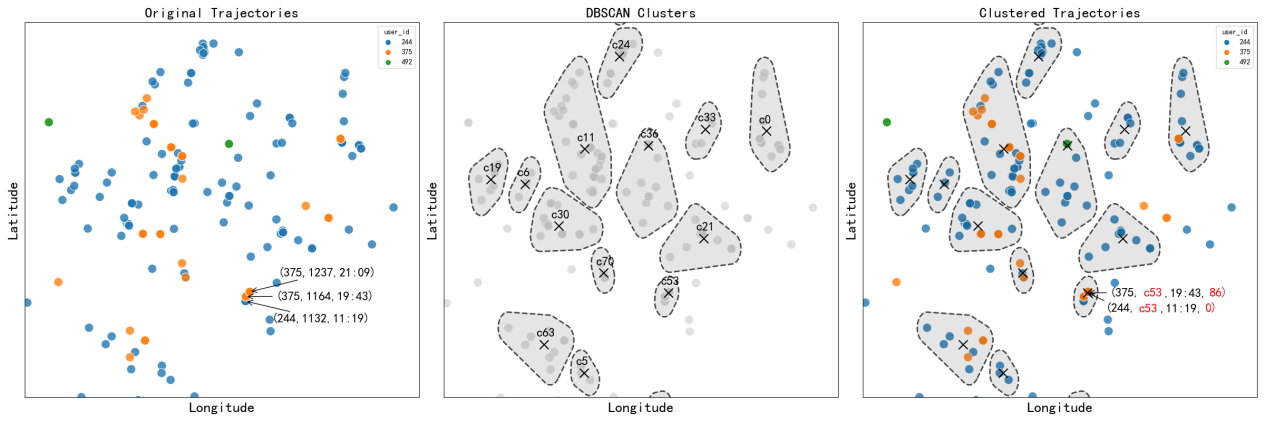}
  \caption{Illustration of Spatio-Temporal Trajectory Preprocessing}
  \label{fig:DBSCANl}
\end{figure*}

\item \textbf{Activity event aggregation and spatiotemporal semantic enrichment.}

After completing spatial discretization, the core task of this step is to perform
aggregation along the temporal dimension in order to endow trajectory data with
spatiotemporal semantics. For each user $u_i$, given the raw trajectory
$\mathrm{tra}(u_i) = \{r_1, r_2, \ldots, r_m\}$, the sequence of check-in points is
traversed. If a temporally consecutive segment of check-ins
$\{r_k, \ldots, r_{k+x}\}$ is mapped to the same semantic anchor $c_j$, this
subsequence is aggregated into a single spatiotemporal event $p_k$.

The event is formalized as a quadruple
$ p_k = (u_i, c_k, t_k, d_k), $
where
\begin{itemize}
  \item $c_k$ denotes the semantic anchor at which the event occurs, represented
        by the cluster centroid coordinates;
  \item $t_k$ is the start time of the event, i.e., the timestamp of the first
        check-in in the subsequence;
  \item $d_k$ is the activity duration of the event, defined as the difference
        between the timestamps of the last and the first check-ins in the
        subsequence, i.e.,
        $ d_k = t_{k+x} - t_k. $
\end{itemize}
In this formulation, the introduction of \emph{activity duration} serves as the
key to semantic elevation. It integrates originally isolated \emph{check-in
points} into \emph{activity periods} with inherent temporal spans, thereby
providing the model with a clear quantitative basis for distinguishing different
behavioral patterns, such as short stops and long stays. Through this step, a
user's raw trajectory is successfully elevated from a low-level sequence of
coordinate points to an activity event sequence composed of spatiotemporal
events,$\mathrm{seq}(u_i) = \{p_1, p_2, \ldots, p_M\}$,
which encodes richer behavioral logic and contextual information.

\item \textbf{Sequence normalization.}
To satisfy the requirement of fixed-length inputs for deep learning models, all
users' activity event sequences are normalized. Specifically, a unified sequence
length $n$ is set as a hyperparameter. For users whose number of activity events
$M$ exceeds $n$, only the most recent $n$ events in time are retained; for users
with $M < n$, zero values are padded at the end of the sequence until the length
reaches $n$. During model training, these padded positions are masked out to
prevent them from affecting the actual computations. After this processing step,
all users' trajectory data are standardized into tensors of uniform
dimensionality,
$\mathrm{seq}(u_i) = \{p_1, p_2, \ldots, p_n\}$,
facilitating their input into subsequent network models.
\end{enumerate}

\subsection{Long- and Short-Term Decoupled Multimodal Embedding}
After obtaining high-quality activity event sequences, this section moves into
the key stage of representation learning, namely transforming these sequences
into differentiated multimodal embeddings suitable for attention-based networks.
The theoretical motivation for this design originates from a classic assertion
in behavioral science, which states that human travel decisions are jointly
driven by long-term stable habits and short-term immediate intentions
\cite{triandis1977behavior}. To effectively capture the complex interactions
between these two patterns, we design a dual-stream architecture.

It is worth noting that the \emph{dual-stream} concept here does not refer to a
physical separation of data sources; rather, its core idea lies in functional
decoupling at the architectural level. Specifically, the long-term pattern stream
aims to distill stable, periodic, and regular patterns from historical behaviors,
while the short-term context stream focuses on capturing highly dynamic
situational intentions driven by recent events. In view of the differentiated
characteristics of these two types of information, corresponding embedding
schemes are further designed, enabling the model to learn these complementary
behavioral drivers in a more specialized manner and to provide more
discriminative input representations for the subsequent dynamic fusion module.
\subsubsection{Time-Window-Based Functional Partitioning of Sequences}

The implementation of the dual-stream architecture first relies on a functional
partitioning of the input activity event sequence. This partitioning logically
defines two subsets, allowing subsequent processing streams to focus on
information at different temporal scales.

Given a user's activity event sequence
$\mathrm{seq}(u_i) = \{p_1, p_2, \ldots, p_n\}$,
where each event is defined as $p_k = (u_i, c_k, t_k, d_k)$, we take the current
prediction time point $t_{\text{predict}}$ as a reference and introduce a
time-window hyperparameter $H_{\text{short}}$ (set to 48 hours by default).
Based on this time window, the event indices in the sequence are divided into two
mutually exclusive sets.

\paragraph{Short-term context index set.}
The short-term context index set, denoted as $I_{\text{short}}$, contains all
event indices that occur within the recent time window. It is formally defined as
$I_{\text{short}} = \{ k \mid (t_{\text{predict}} - t_k) \le H_{\text{short}} \}$.
The corresponding subsequence
$S_{\text{short}} = \{ p_k \mid k \in I_{\text{short}} \}$
mainly reflects the user's immediate state and situational intentions.

\paragraph{Long-term pattern index set.}
The long-term pattern index set, denoted as $I_{\text{long}}$, consists of the
remaining, relatively earlier event indices in the input sequence:
$I_{\text{long}} = \{ k \mid (t_{\text{predict}} - t_k) > H_{\text{short}} \}$.
The corresponding subsequence
$S_{\text{long}} = \{ p_k \mid k \in I_{\text{long}} \}$
represents a baseline of the user's more stable and habitual behavioral patterns.

In the model implementation, both parallel processing streams receive the
complete input sequence tensor. Based on the above index sets, two binary mask
matrices, $\mathbf{M}_{\text{short}}$ and $\mathbf{M}_{\text{long}}$, are
generated. By applying the corresponding masks in their respective attention
computations, the model is guided to functionally separate the processing of
long-term and short-term information without altering the underlying data
structure, thereby laying the foundation for differentiated embedding learning
and feature extraction.
\subsubsection{Differentiated Multimodal Embedding under the Dual-Stream Architecture}

After completing the functional partitioning of the activity event sequence, this
section further introduces a differentiated embedding mechanism. This mechanism
consists of two components, namely \emph{activity event embeddings} and
\emph{spatiotemporal relational embeddings}, which model information
characteristics from different dimensions. Together, they strengthen the
functional positioning and complementary roles of the long- and short-term
processing streams at the feature level.

\paragraph{Activity event embeddings.}
Activity event embeddings aim to map each discrete spatiotemporal event
$p_k = (u_i, c_k, t_k, d_k)$ into a high-dimensional vector representation. While
the embedding schemes of the long- and short-term streams share several basic
components, they are differentiated along key dimensions according to their
respective modeling objectives.

\paragraph{Short-term context stream.}
For the short-term context stream, the embedding representation focuses on event
immediacy and high-precision temporal ordering. The composite embedding of event
$p_k$ in this stream, denoted as $\mathbf{e}^{\text{short}}_k$, is constructed as
a linear combination of the following components.

\emph{Base embeddings.}
This component includes the user embedding
$\mathbf{E}^{U}_{u_i} \in \mathbb{R}^d$ and the location embedding
$\mathbf{E}^{L}_{c_k} \in \mathbb{R}^d$, which are learned via standard embedding
layers to capture the static characteristics of users and locations.

\emph{High-precision temporal embedding.}
To capture precise temporal relationships among recent events, sinusoidal and
cosine functions are employed to perform positional encoding on the continuous
timestamp $t_k$. This design preserves high-frequency temporal information and is
crucial for distinguishing consecutive activities separated by minutes or hours.

\emph{Activity duration embedding.}
The continuous activity duration $d_k$ is first discretized into predefined
intervals (e.g., ``$<$5 minutes,'' ``5--30 minutes,'' ``30--120 minutes,'' and
``$>$120 minutes''), and an embedding vector
$\mathbf{E}^{D}_{d_k} \in \mathbb{R}^d$ is learned for each interval. The
introduction of this embedding constitutes a key enhancement of the proposed
model, as it explicitly provides quantitative cues for distinguishing short stays
(e.g., transfers) from long stays (e.g., work), thereby significantly enriching
the semantic content of activity events.

The resulting composite embedding is given by
\begin{equation}
\mathbf{e}^{\text{short}}_k =
\mathbf{E}^{U}_{u_i} +
\mathbf{E}^{L}_{c_k} +
\mathbf{E}^{T}_{t_k} +
\mathbf{E}^{D}_{d_k}.
\label{eq:short_event_embedding}
\end{equation}
Finally, the short-term context event sequence is jointly represented as an
embedding matrix
$\mathbf{E}^{\text{short}}_{u_i}$
= $\{\mathbf{e}^{\text{short}}_1, \mathbf{e}^{\text{short}}_2, \ldots,
\mathbf{e}^{\text{short}}_n\}
\in \mathbb{R}^{n \times d}$.

\paragraph{Long-term pattern stream.}
For the long-term pattern stream, the core objective of its embedding
representation is to capture behavioral periodicity and macroscopic regularities.
While sharing the base embeddings, the event embedding
$\mathbf{e}^{\text{long}}_k$ is specifically designed with respect to
time-related features.

\emph{Periodic temporal embedding.}
Unlike the short-term stream, the temporal embedding in this stream focuses on
macroscopic cycles. The continuous timestamp $t_k$ is discretized into specific
time slots within a week (a total of $7 \times 24 = 168$ categories), and an
embedding vector is learned for each slot. This design explicitly captures the
weekly cyclic rhythm of user behavior.

\emph{Macroscopic periodic feature embedding.}
To further strengthen the modeling of stable habits, this stream additionally
introduces a macroscopic periodic feature embedding
$\mathbf{E}^{P}_{\text{period}}(t_k)$. Two discrete features, namely the day of
the week and the time of day, are extracted from the timestamp $t_k$, and
independent embedding vectors are learned for them. This design explicitly
endows the model with the ability to recognize macroscopic contexts such as
\emph{weekday morning rush hours} or \emph{weekend evenings}, enabling it to more
effectively distill users' core behavioral regularities.

The composite embedding for the long-term stream is expressed as
\begin{equation}
\mathbf{e}^{\text{long}}_k =
\mathbf{E}^{U}_{u_i} +
\mathbf{E}^{L}_{c_k} +
\mathbf{E}^{T}_{t_k} +
\mathbf{E}^{D}_{d_k} +
\mathbf{E}^{P}_{\text{period}}(t_k).
\label{eq:long_event_embedding}
\end{equation}

Accordingly, the long-term historical event sequence is represented as an
embedding matrix
$\mathbf{E}^{\text{long}}_{u_i}$
= $\{\mathbf{e}^{\text{long}}_1, \mathbf{e}^{\text{long}}_2, \ldots,
\mathbf{e}^{\text{long}}_n\}
\in \mathbb{R}^{n \times d}$.

\paragraph{Spatiotemporal relational embeddings.}
Spatiotemporal relational embeddings aim to capture relative relationships
between events, which are equally critical for understanding behavioral logic
beyond the intrinsic attributes of individual events. To avoid the feature
sparsity caused by directly discretizing continuous spatiotemporal intervals,
we adopt the Unit Embedding Layer proposed in STAN \cite{luo2021stan} to generate
dense representations for spatiotemporal intervals.

A key improvement of this module lies in the design of independent unit
embeddings for the long- and short-term streams, enabling the model to
distinguish relational patterns at different temporal scales and thereby more
robustly integrate long-term regularities with short-term intentions.

Specifically, for each stream $\text{stream} \in \{\text{short}, \text{long}\}$,
we construct a spatiotemporal relation matrix within the trajectory
\[
\boldsymbol{\Delta}^{\text{stream}}_t \in \mathbb{R}^{n \times n},
\qquad
\boldsymbol{\Delta}^{\text{stream}}_s \in \mathbb{R}^{n \times n}.
\]
and a spatiotemporal relation matrix between the trajectory and candidate
locations
\[
\mathbf{N}^{\text{stream}}_{t,s} \in \mathbb{R}^{L \times n}.
\]

\emph{Short-term stream unit embeddings.}
The unit embeddings
$(\mathbf{e}^{\text{short}}_{\Delta t}, \mathbf{e}^{\text{short}}_{\Delta s})$
are trained to capture fine-grained spatiotemporal dependencies, sensitively
reflecting the influence of temporal intervals on the order of minutes or hours
and short-range spatial proximity.

\emph{Long-term stream unit embeddings.}
The unit embeddings
$(\mathbf{e}^{\text{long}}_{\Delta t}, \mathbf{e}^{\text{long}}_{\Delta s})$
focus on macroscopic and periodic relationships, emphasizing cyclically
meaningful intervals such as 24 hours (daily) or 168 hours (weekly), in service
of modeling stable behavioral habits.

Through the above process, the original spatiotemporal interval matrices are
transformed into corresponding embedding tensors:
\[
\mathbf{E}^{\text{stream}}_{\Delta t}, \mathbf{E}^{\text{stream}}_{\Delta s}
\in \mathbb{R}^{n \times n \times d},
\qquad
\mathbf{E}^{\text{stream}}_{N t}, \mathbf{E}^{\text{stream}}_{N s}
\in \mathbb{R}^{L \times n \times d}.
\]

The temporal and spatial components are then combined via weighted summation to
obtain the final composite relational embeddings:
\begin{align}
\mathbf{E}^{\text{stream}}_{\Delta}
&= \mathrm{Sum}\!\left(\mathbf{E}^{\text{stream}}_{\Delta t}\right)
 + \mathrm{Sum}\!\left(\mathbf{E}^{\text{stream}}_{\Delta s}\right),
\label{eq:relation_traj}\\
\mathbf{E}^{\text{stream}}_{N}
&= \mathrm{Sum}\!\left(\mathbf{E}^{\text{stream}}_{N t}\right)
 + \mathrm{Sum}\!\left(\mathbf{E}^{\text{stream}}_{N s}\right),
\label{eq:relation_candidate}
\end{align}
where $\text{stream} \in \{\text{short}, \text{long}\}$.

At this point, the dual-stream architecture has constructed comprehensive and
differentiated multimodal embedding representations, encompassing both
fine-grained characterizations of event attributes and scale-aware modeling of
spatiotemporal structural relationships. These information-rich and functionally
specialized representations provide high-quality inputs for the subsequent
spatiotemporal attention network.

\subsection{Environment-Feature-Enhanced Spatio-Temporal Attention Network}

This section introduces the final stage of the model framework, namely the
environment-feature-enhanced spatiotemporal attention network. As the
decision-making core, this network is designed to fuse the differentiated
long-term and short-term embeddings obtained in the previous stage, while
explicitly incorporating external contextual information such as weather
conditions and real-time traffic states. Simple feature concatenation or static
weighting schemes ignore the dynamic and context-dependent nature of user
decision-making, which would weaken the benefits brought by prior feature
decoupling and limit the representational capacity of the model. To address this
issue, the proposed network consists of three key components:

\begin{enumerate}[label=(\arabic*), leftmargin=*]
\item a dual-stream self-attention aggregation module, which separately models
the internal dependency structures of long-term and short-term embeddings;
\item a context-aware dynamic fusion module, which adaptively allocates the
relative influence of the two streams according to the current scenario and
external environmental features; and
\item an attention-based matching and prediction module, which computes
probability scores over all candidate locations.
\end{enumerate}

By introducing and leveraging environmental features during the fusion stage,
the model can dynamically balance long-term and short-term driving factors at
the sample level, thereby enabling fine-grained modeling of individual travel
decision processes and improving both prediction performance and
interpretability.

\subsubsection{Dual-Stream Self-Attention Aggregation}

This module exploits the ability of the self-attention mechanism to break
sequential order constraints and directly capture relationships between any two
activity events. Two parallel processing streams are designed to aggregate
long-term and short-term embedding representations separately, thereby deeply
mining the spatiotemporal dependencies within the respective subsequences.

The core of this mechanism is scaled dot-product attention augmented with
spatiotemporal bias. For each processing stream
$\text{stream} \in \{\text{short}, \text{long}\}$, the inputs include the activity
event embeddings $\mathbf{E}^{\text{stream}}_{u_i}$ and the spatiotemporal
relational embeddings $\mathbf{E}^{\text{stream}}_{\Delta}$. First, the activity
event embeddings are projected through three stream-specific linear
transformations to generate the query, key, and value matrices, i.e.,
$\mathbf{Q}^{\text{stream}} = \mathbf{E}^{\text{stream}}_{u_i}\mathbf{W}^{Q}_{\text{stream}}$,
with $\mathbf{K}^{\text{stream}}$ and $\mathbf{V}^{\text{stream}}$ defined
analogously.

The key innovation of this design lies in directly injecting the spatiotemporal
relational embedding $\mathbf{E}^{\text{stream}}_{\Delta}$ as a bias term into
the similarity computation between queries and keys. This mechanism forces the
model, when evaluating the relevance between any two activity events, to jointly
consider their intrinsic semantic similarity (captured by the dot product
$\mathbf{Q}\mathbf{K}^\top$) and their spatiotemporal proximity or periodic
relationships encoded by the relational embeddings. By further applying the
corresponding binary mask matrix $\mathbf{M}_{\text{stream}}$ and Softmax
normalization, the attention weights are obtained and used to produce a new
sequence representation:

\begin{equation}
\scriptsize
\mathbf{R}^{\text{stream}}
=
\Big(
\mathbf{M}_{\text{stream}}
\odot
\mathrm{Softmax}\Big(
\mathbf{Q}^{\text{stream}}
\mathbf{K}^{\text{stream}\top}
+
\frac{\mathbf{E}^{\text{stream}}_{\Delta}}{\sqrt{d}}
\Big)
\Big)
\mathbf{V}^{\text{stream}} .
\label{eq:dual_stream_attention}
\end{equation}

Through this parallel dual-stream processing, the model is able to simultaneously
and independently learn two types of dependencies with distinct characteristics.
The long-term pattern stream, whose inputs emphasize periodicity and historical
regularity, tends to associate temporally distant events with stable behavioral
patterns, thereby distilling the user’s core habits. In contrast, the short-term
context stream focuses on high-resolution temporal information, making it more
sensitive to recently occurring and logically connected activity sequences.
Ultimately, this module outputs two functionally specialized representations,
$\mathbf{R}^{\text{long}}$ and $\mathbf{R}^{\text{short}}$, which jointly provide
semantically richer and more discriminative inputs for downstream prediction.

\subsubsection{Context-Aware Dynamic Fusion}

After obtaining feature representations that separately characterize the
user’s stable habits $\mathbf{R}^{\text{long}}$ and immediate intentions
$\mathbf{R}^{\text{short}}$ through self-attention aggregation, the goal of this
module is to intelligently and personalizedly fuse these two decision drivers
according to the current context. Static or simplistic fusion strategies (e.g.,
concatenation or averaging) fail to adapt to the varying decision logic across
different scenarios, since the relative importance of long-term habits and
short-term intentions is not fixed but dynamically changes with the external
environment, user state, and temporal context. To this end, we design a
gating-mechanism-based dynamic fusion module, whose fusion weights are generated
adaptively from a carefully constructed context vector.

\paragraph{External context modeling.}
To enable the model to perceive the objective environment in which the prediction
task takes place, heterogeneous external knowledge sources are first unified into
a common representation. Specifically, the model incorporates three types of
external information that are closely related to travel decisions: (1) origin
POI type, which describes the semantic attributes of candidate locations (e.g.,
\emph{restaurant} or \emph{transport hub}) and provides prior knowledge about
destination characteristics; (2) traffic congestion index, which reflects
real-time traffic conditions and directly affects travel feasibility and time
cost; and (3) weather conditions, an important environmental factor influencing
travel willingness and mode choice.

These external features, whether discrete categories (e.g., POI type and weather
conditions) or continuous values (e.g., congestion index), are first
appropriately discretized and then mapped into dense vector representations via
their respective embedding layers. The resulting knowledge embeddings are
concatenated and passed through a lightweight feedforward network for information
fusion, yielding a unified external knowledge embedding vector
$\mathbf{e}_{\text{ext}}$. This vector can be interpreted as a comprehensive
snapshot of the objective environment, answering the question: \emph{where can I
go now, and how convenient is it to go there?}

\paragraph{Decision context construction.}
A complete decision context should capture not only the external environment, but
also the user’s personalized preferences and the current temporal background,
since identical external conditions may lead to drastically different decisions
for different users or at different times. Accordingly, the final context vector
$\mathbf{C}$ is constructed by concatenating three key components that correspond
to the core questions in the decision-making process: \emph{How is the
environment?}, \emph{Who am I?}, and \emph{What time is it now?}. Specifically,
$\mathbf{C}$ consists of the external knowledge embedding
$\mathbf{e}_{\text{ext}}$, the user embedding $\mathbf{E}^{U}_{u_i}$, and the
current high-resolution temporal embedding $\mathbf{E}^{T}_{t_{\text{predict}}}$.

The external knowledge embedding provides objective environmental constraints;
the user embedding introduces a personalized perspective, ensuring that fusion
weights vary across users (e.g., severe traffic congestion may significantly
affect users who usually drive but have little impact on those who prefer public
transit); and the temporal embedding supplies contextual cues that allow the
model to recognize indicative time patterns such as \emph{weekday morning rush
hours} or \emph{weekend evenings}, thereby facilitating the activation of
relevant historical habits.

\paragraph{Dynamic gating mechanism.}
The context vector $\mathbf{C}$ is fed into a gating network, implemented as a
feedforward neural network with a Sigmoid activation function $\sigma(\cdot)$.
Its objective is to compute a dynamic gating weight
$g \in [0,1]$ for the long-term pattern stream based on the current decision
context:
\begin{equation}
g = \sigma\!\left(
\mathbf{W}_g
\left[
\mathbf{e}_{\text{ext}} \,;\,
\mathbf{E}^{U}_{u_i} \,;\,
\mathbf{E}^{T}_{t_{\text{predict}}}
\right]
+ \mathbf{b}_g
\right).
\label{eq:dynamic_gating}
\end{equation}
This gating weight $g$ can be intuitively interpreted as the degree to which the
user’s decision under the current scenario relies on long-term stable habits.
Correspondingly, $1 - g$ represents the degree of reliance on recent events and
immediate intentions. The final fused feature representation
$\mathbf{R}^{\text{fused}}$ is obtained by a weighted combination of the long-
and short-term representations using this dynamic gating weight:
\begin{equation}
\mathbf{R}^{\text{fused}}
=
g \cdot \mathbf{R}^{\text{long}}
+
(1 - g) \cdot \mathbf{R}^{\text{short}}.
\label{eq:fusion_representation}
\end{equation}

With this design, the model can adaptively adjust its fusion strategy for each
prediction instance based on rich contextual information. For example, when
external knowledge indicates poor weather and heavy traffic, and both user and
time embeddings correspond to a typical commuting scenario, the learned gating
weight $g$ naturally approaches $1$, amplifying the influence of long-term
commuting patterns $\mathbf{R}^{\text{long}}$. Conversely, when the user is in a
new commercial area during an irregular weekend time slot, $g$ tends toward $0$,
assigning greater importance to the short-term context
$\mathbf{R}^{\text{short}}$, which is more effective at capturing exploratory
and opportunistic behaviors. This dynamic, personalized, and context-aware
fusion mechanism provides a fine-grained simulation of complex user decision
processes and constitutes a key factor in improving both predictive accuracy and
model interpretability.

\subsubsection{Attention-Based Matching and Probability Output}

After obtaining the fused comprehensive feature representation
$\mathbf{R}^{\text{fused}}$, the model enters the final prediction stage, whose
objective is to compute a probability distribution over all candidate locations.
To this end, an attention-based matching layer inspired by STAN
\cite{luo2021stan} is employed, which incorporates all historical trajectory
information into the matching process to fully exploit rich spatiotemporal
details.

This matching procedure can be regarded as a spatiotemporal-aware similarity
computation. Specifically, the embeddings of candidate locations
$\mathbf{E}^{L}$ are used as the query matrix $\mathbf{Q}^{\text{match}}$, while
the fused trajectory representation $\mathbf{R}^{\text{fused}}$ serves as the
key matrix $\mathbf{K}^{\text{match}}$. During similarity computation, the
dynamically fused spatiotemporal relational embedding
$\mathbf{E}^{\text{fused}}_{N}$ is injected as a bias term, which is obtained as
a weighted combination of the long- and short-term relational embeddings,
i.e.,
$\mathbf{E}^{\text{fused}}_{N}
=
g \cdot \mathbf{E}^{\text{long}}_{N}
+
(1 - g) \cdot \mathbf{E}^{\text{short}}_{N}$.

Finally, the probability score of each candidate location is computed through
the attention mechanism. The Softmax function is applied row-wise to the
similarity matrix to obtain attention weights over all historical events, and
these weights are subsequently aggregated to produce the final probability
vector $\mathbf{P}_{u_i}$:
\begin{equation}
\mathbf{P}_{u_i}
=
\mathrm{Sum}\!\left(
\mathrm{Softmax}\!\left(
\mathbf{Q}^{\text{match}}
\mathbf{K}^{\text{match}\top}
+
\mathbf{E}^{\text{fused}}_{N} / \sqrt{d}
\right)
\right).
\label{eq:prediction_probability}
\end{equation}

The summation operation ensures that every historical event contributes to the
prediction outcome, thereby explicitly preserving repeat-visit frequency
patterns and enhancing the personalized expressiveness of the prediction.
During model training, a cross-entropy loss with negative sampling is employed
for optimization, enabling efficient and accurate learning in the presence of
large numbers of negative candidate locations.

\section{Experimental Results Analysis}

\subsection{Datasets}

We conduct experiments on two real-world datasets, namely Foursquare-NYC and
Foursquare-TKY. Both datasets are derived from the Foursquare platform and
contain user trajectory data collected over an 11-month period in New York City
and Tokyo, respectively.

The preprocessing procedure is as follows. First, check-in records of each user
are sorted chronologically, and low-frequency points of interest (POIs) with
fewer than five visits are removed. Next, check-in records are partitioned into
sessions using a 24-hour time window; sessions containing fewer than three
check-ins are discarded, and inactive users are filtered out. For each user with
$m$ check-in records, we split the dataset into training and test sets by
assigning the first 80\% of sessions to the training set and the remaining 20\%
to the test set.

\subsection{Evaluation Metrics}

To evaluate model performance, we adopt two widely used metrics: Recall@K and
Normalized Discounted Cumulative Gain (NDCG@K), where $K$ is set to 5 and 10.

Recall@K measures whether the ground-truth item appears within the top-$K$
recommended results. NDCG@K evaluates the ranking quality of the recommendation
list by computing the ratio between Discounted Cumulative Gain (DCG) and Ideal
Discounted Cumulative Gain (IDCG), thereby reflecting the effectiveness of the
model in ranking relevant items at higher positions.

In our experiments, we report NDCG@5 and NDCG@10 to provide a comprehensive
assessment of both recommendation accuracy and ranking quality.

\subsection{Performance Comparison}

To evaluate the effectiveness of the proposed model in the next-item
recommendation task, we select representative baseline methods from five
different categories, including statistical models, recurrent neural networks
(RNNs), self-attention-based models, graph neural networks (GNNs), and
contrastive learning frameworks.
\paragraph{Statistical Methods.}
\begin{itemize}
  \item \textbf{UserPop}: Recommends the most frequently visited POIs based on a
  user’s historical visiting frequency. This method serves as a fundamental
  statistical baseline.
\end{itemize}

\paragraph{RNN-based Methods.}
\begin{itemize}
  \item \textbf{STGN} \cite{zhao2020stgn}: Incorporates spatial and temporal gates
  into the LSTM architecture to jointly model long-term and short-term user
  preferences.
  \item \textbf{LSTPM} \cite{sun2020lstpm}: Integrates a non-local mechanism and a
  geographical extension module into LSTM to enhance the modeling of trajectory
  features.
\end{itemize}

\paragraph{Self-Attention-based Methods.}
\begin{itemize}
  \item \textbf{STAN} \cite{luo2021stan}: Utilizes self-attention networks to
  explicitly model temporal and spatial features in check-in sequences, enabling
  the capture of complex trajectory dependencies.
\end{itemize}

\paragraph{GNN and Hypergraph-based Methods.}
\begin{itemize}
  \item \textbf{LightGCN} \cite{he2020lightgcn}: Implements efficient
  collaborative filtering by simplifying graph convolution operations and
  removing non-linear transformations.
  \item \textbf{SGRec} \cite{li2021sgrec}: Models local collaborative signals
  through sequence-augmented graphs.
  \item \textbf{GETNext} \cite{yang2022getnext}: Combines graph neural networks
  and Transformer architectures to fuse global transition patterns with
  collaborative features.
  \item \textbf{MSTHN} \cite{lai2023msthn}: Employs a multi-view spatio-temporal
  hypergraph to jointly model high-order collaborative relationships between
  users and POIs.
  \item \textbf{STHGCN} \cite{yan2023sthgcn}: Focuses on spatio-temporal
  relationships among complex trajectories and extracts global collaborative
  features via hypergraph modeling.
  \item \textbf{DCHL} \cite{lai2024dchl}: Proposes a disentangled contrastive
  hypergraph learning framework to dynamically model multi-dimensional user
  preferences and cross-view interactions.
\end{itemize}

\paragraph{Graph/Hypergraph Contrastive Learning Methods.}
\begin{itemize}
  \item \textbf{DisenPOI} \cite{qin2023disenpoi}: Utilizes graph contrastive
  learning to disentangle geographical and sequential influences, thereby
  enhancing representation capability.
  \item \textbf{HCCF} \cite{xia2022hccf}: A hypergraph-enhanced cross-view
  contrastive learning framework that fuses both local and global collaborative
  information for recommendation.
\end{itemize}

To ensure a fair comparison, category information for POIs in SGRec, GETNext,
and STHGCN is removed during the experiments, so that all methods are evaluated
under a consistent feature setting.

Results Analysis

We performed a systematic comparative analysis of the proposed model and multiple baselines on the TKY and NYC datasets, with the results summarized in Table~\ref{tab:performance}. Significant performance variances across models can be observed for Recall@5, Recall@10, NDCG@5, and NDCG@10.

Traditional statistical methods and simple sequential models exhibit overall poor performance. Specifically, UserPop, STGN, and LSTPM underperform across multiple metrics (e.g., UserPop achieves an NDCG@10 of only 0.1861 on the TKY dataset). This indicates that such methods struggle to effectively characterize complex spatio-temporal dependency structures.

In contrast, STAN achieves performance gains over traditional models via the self-attention mechanism; however, it fails to fully utilize critical contextual information such as activity duration, leading to limitations in modeling precision and generalization. Among GNN-based methods, SGRec and GETNext consistently outperform LightGCN on both datasets, validating the necessity of integrating spatio-temporal transition patterns with collaborative signals.

Furthermore, hypergraph-based models (such as DCHL, MSTHN, and STHGCN) achieve Recall@10 scores of 0.4083, 0.3927, and 0.3924 on the TKY dataset, respectively—representing an approximate 20\% improvement over GNN-based methods. This result underscores the effectiveness of hypergraph structures in modeling high-order collaborative relationships. While contrastive learning methods such as DisenPOI and HCCF also demonstrate improvements, their overall performance remains suboptimal due to insufficient multi-view information fusion and incomplete decoupling of latent representations.

In summary, the experimental results demonstrate that deep modeling of spatio-temporal dependencies combined with rich contextual semantic features constitutes a key direction for further enhancing next-location recommendation performance.
\begin{table*}[t]
\centering
\caption{Performance comparison on TKY and NYC datasets}
\label{tab:performance}
\resizebox{\textwidth}{!}{
\begin{tabular}{l|cccc|cccc}
\hline
\multirow{2}{*}{Method} & \multicolumn{4}{c|}{TKY} & \multicolumn{4}{c}{NYC} \\
& Recall@5 & Recall@10 & NDCG@5 & NDCG@10
& Recall@5 & Recall@10 & NDCG@5 & NDCG@10 \\
\hline
UserPop   & 0.2229 & 0.2668 & 0.1718 & 0.1861 & 0.2866 & 0.3297 & 0.2283 & 0.2423 \\
STGN      & 0.2112 & 0.2587 & 0.1482 & 0.1589 & 0.2371 & 0.2594 & 0.2261 & 0.2307 \\
LSTPM     & 0.2203 & 0.2703 & 0.1556 & 0.1734 & 0.2495 & 0.2668 & 0.2425 & 0.2483 \\
\hline
STAN      & 0.2621 & 0.3317 & 0.2074 & 0.2189 & 0.3523 & 0.3827 & 0.3025 & 0.3137 \\
LightGCN  & 0.2213 & 0.2594 & 0.1977 & 0.2098 & 0.3221 & 0.3488 & 0.2958 & 0.3042 \\
SGRec     & 0.2537 & 0.3213 & 0.2221 & 0.2447 & 0.3451 & 0.3723 & 0.3052 & 0.3178 \\
GETNext   & 0.2686 & 0.3282 & 0.2212 & 0.2242 & 0.3572 & 0.3866 & 0.3079 & 0.3094 \\
\hline
MSTHN     & 0.3378 & 0.3927 & 0.2567 & 0.2721 & 0.4076 & 0.4398 & 0.3612 & 0.3702 \\
STHGCN    & 0.3392 & 0.3924 & 0.2592 & 0.2693 & 0.4081 & 0.4366 & 0.3626 & 0.3703 \\
DisenPOI  & 0.2692 & 0.3314 & 0.2263 & 0.2332 & 0.3577 & 0.3831 & 0.2979 & 0.3071 \\
HCCF      & 0.2689 & 0.3253 & 0.2325 & 0.2429 & 0.3534 & 0.3745 & 0.3025 & 0.3134 \\
DCHL      & 0.3662 & 0.4083 & 0.2951 & 0.3078 & 0.4385 & 0.4861 & 0.3859 & 0.4017 \\
\hline
Ours      & \textbf{0.4400} & \textbf{0.6200} & \textbf{0.2972} & \textbf{0.3556}
          & \textbf{0.5300} & \textbf{0.6600} & \textbf{0.3943} & \textbf{0.4356} \\
\hline
\end{tabular}
}
\end{table*}

\begin{figure}[t]
  \centering
  \includegraphics[width=\columnwidth]{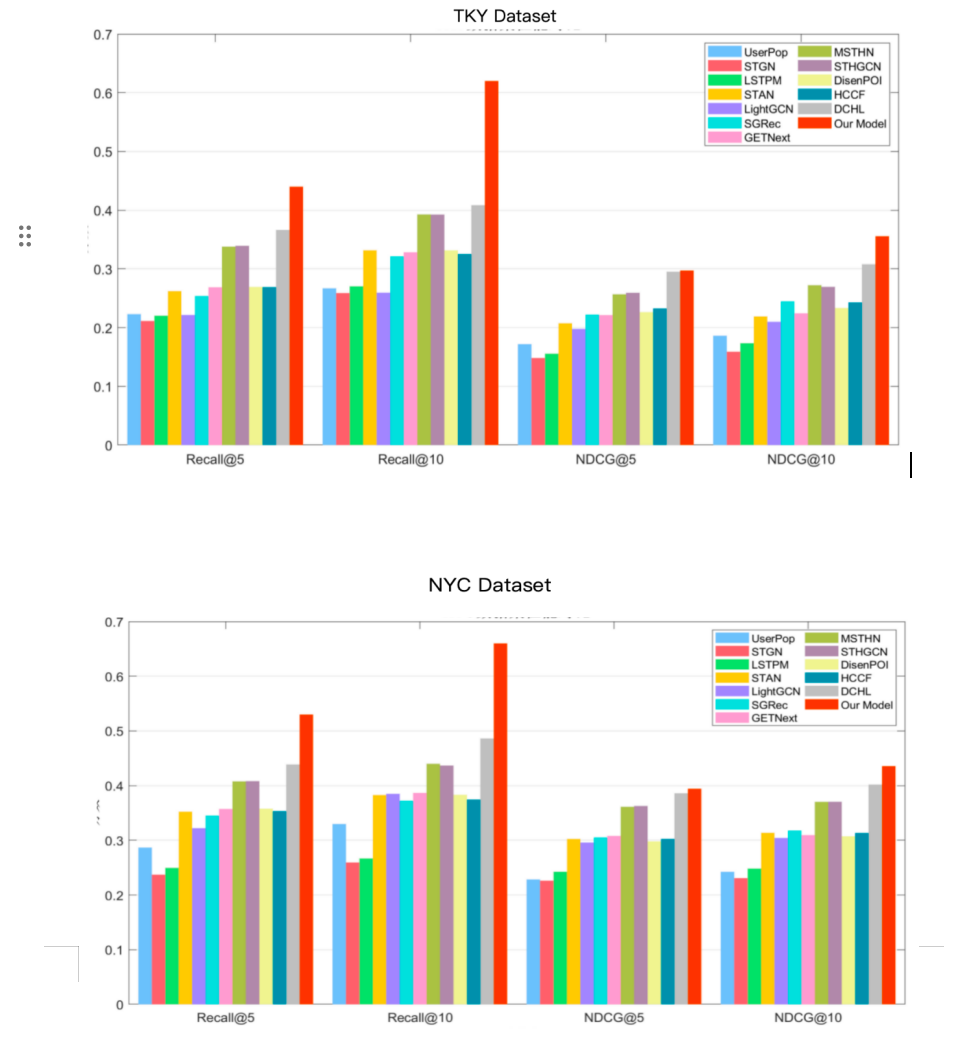}
  \caption{Performance Comparison of Different Models on Two Datasets.}
  \label{fig:performance}
\end{figure}

As illustrated in Figure~\ref{fig:performance}, the proposed model achieves state-of-the-art performance across most evaluation metrics. In particular, the NDCG@5 scores reach 0.2972 and 0.3943 on the TKY and NYC datasets, respectively, demonstrating strong and robust predictive capability. These performance gains can be primarily attributed to the following design considerations:

\begin{itemize}
    \item The introduction of activity duration as a spatio-temporal contextual feature, combined with a dual-layer attention mechanism to enrich trajectory representations;
    \item Comprehensive modeling of long- and short-term user preferences, enabling a deeper understanding of user behavioral patterns;
    \item Explicit integration of temporal features during the trajectory embedding phase, which strengthens the modeling of behavioral dependencies;
    \item Incorporation of temporal factors during the attention-based matching stage, enhancing the precision of candidate location evaluation.
\end{itemize}

\subsection{Ablation Study}

\begin{figure*}[t]
  \centering
  \includegraphics[width=0.95\textwidth]{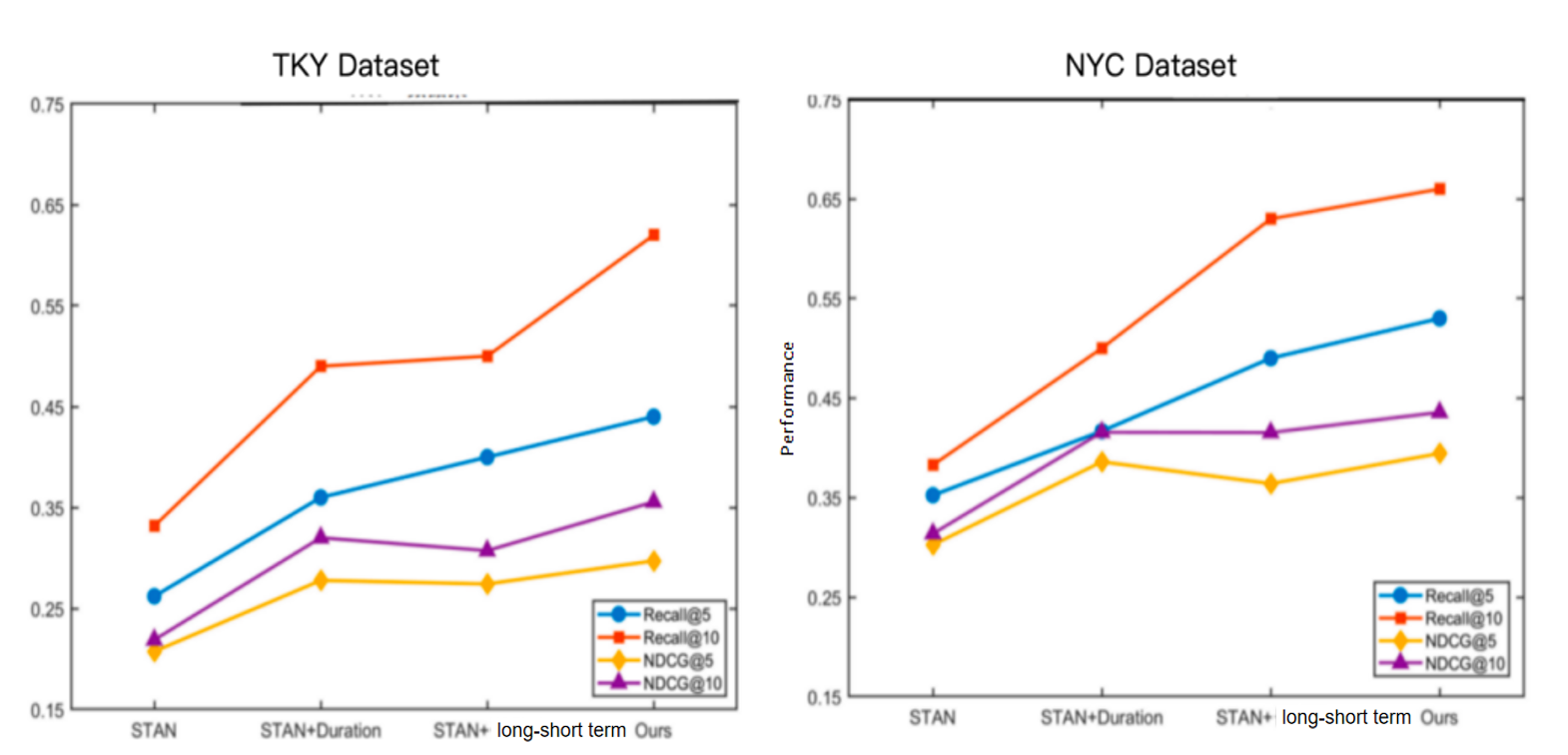}
  \caption{Ablation Study of the Proposed Model}
  \label{fig:ablation}
\end{figure*}

We conducted ablation experiments to analyze the impact of spatio-temporal contextual features, such as activity duration, on model performance. Using STAN as the baseline model, we systematically investigated the effects of different input feature combinations and long- and short-term preference learning mechanisms. Table~\ref{tab:ablation} and Figure~\ref{fig:ablation} report the experimental results in terms of Recall@5, Recall@10, NDCG@5, and NDCG@10.

Although the STAN model demonstrates competitive performance in next-location recommendation, it still exhibits limitations in modeling rich spatio-temporal context. After introducing activity duration features, the model achieves consistent and significant improvements across all evaluation metrics, indicating that activity duration effectively captures users’ behavioral rhythms and activity characteristics, thereby enhancing recommendation accuracy.

Furthermore, incorporating long- and short-term preference learning results in a substantial performance gain, especially in Recall@10, which reaches 0.6300 on the NYC dataset. This observation suggests that explicitly modeling long-term dependencies together with short-term dynamics plays a critical role in understanding user mobility behavior. The complete model outperforms all ablated variants, validating the synergistic effect of integrating activity duration features with long- and short-term preference learning.

Overall, these results demonstrate that deeply mining spatio-temporal contextual information and appropriately modeling preference dynamics are crucial for improving both the accuracy and robustness of next-location recommendation models.

\begin{table*}[t]
\centering
\caption{Ablation Study of the Proposed Model}
\label{tab:ablation}
\resizebox{\textwidth}{!}{
\begin{tabular}{l|cccc|cccc}
\hline
\multirow{2}{*}{Method} & \multicolumn{4}{c|}{TKY} & \multicolumn{4}{c}{NYC} \\
 & Recall@5 & Recall@10 & NDCG@5 & NDCG@10
 & Recall@5 & Recall@10 & NDCG@5 & NDCG@10 \\
\hline
STAN              & 0.2621 & 0.3317 & 0.2074 & 0.2189 & 0.3523 & 0.3827 & 0.3025 & 0.3137 \\
STAN + Duration   & 0.3600 & 0.4900 & 0.2778 & 0.3201 & 0.4167 & 0.5000 & 0.3859 & 0.4156 \\
STAN + L/S Pref.  & 0.4000 & 0.5000 & 0.2743 & 0.3073 & 0.4900 & 0.6300 & 0.3640 & 0.4153 \\
\hline
Ours              & \textbf{0.4400} & \textbf{0.6200} & \textbf{0.2972} & \textbf{0.3556}
                  & \textbf{0.5300} & \textbf{0.6600} & \textbf{0.3943} & \textbf{0.4356} \\
\hline
\end{tabular}
}
\end{table*}

\subsection{Impact of Input Sequence Length on Model Performance}

To systematically evaluate the impact of input sequence length on model performance, we design a controlled experiment. Specifically, we vary the time span of users’ historical trajectories on the TKY and NYC datasets while keeping all other experimental settings unchanged. In this study, the number of check-in records $n$ (as defined in Section~3.1, \emph{Semantically Enhanced Spatio-Temporal Trajectory Preprocessing}) is used to measure the input sequence length.

\begin{figure}[t]
  \centering
  \includegraphics[width=\columnwidth]{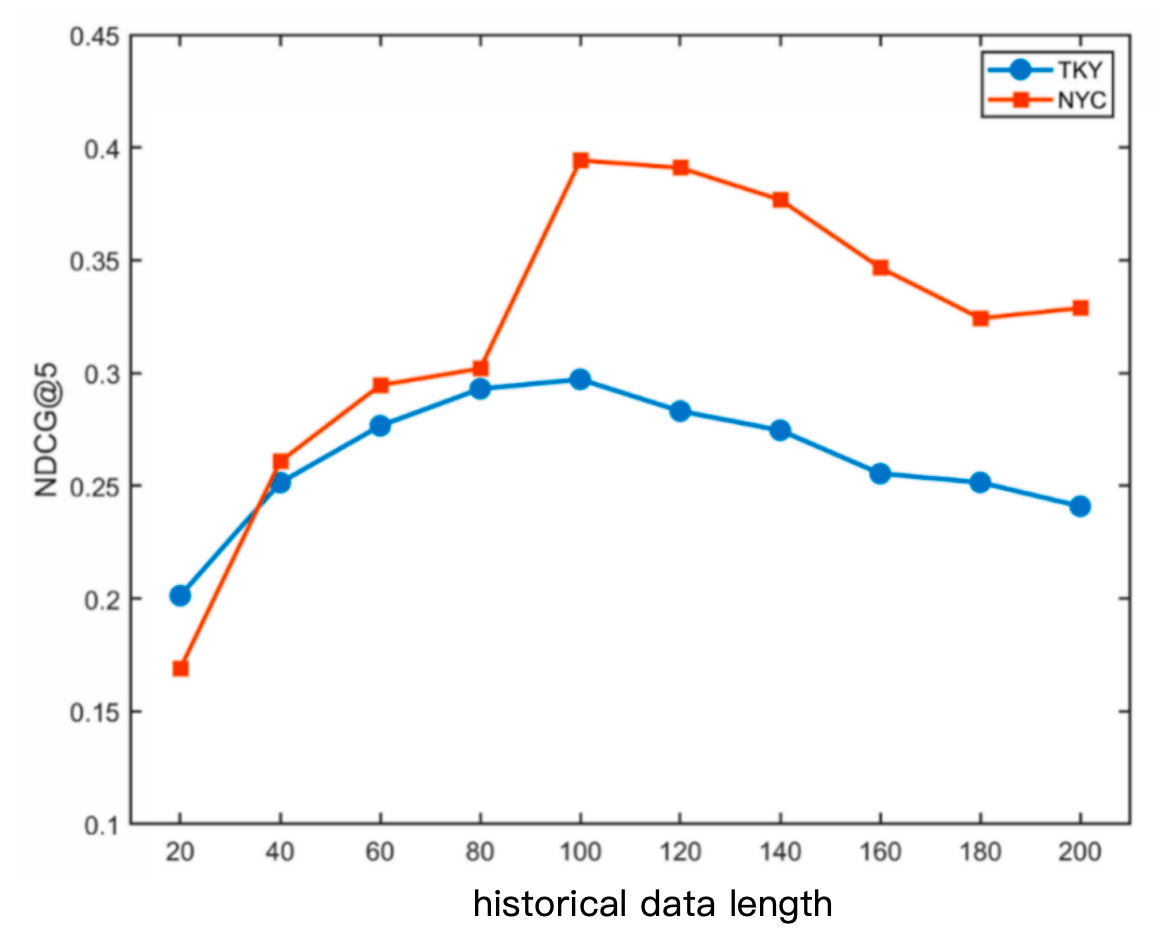}
  \caption{Performance Comparison under Different Historical Input Lengths on Two Datasets.}
  \label{fig:input_length}
\end{figure}

The value of $n$ ranges from 20 to 200 with a step size of 20, resulting in ten different input configurations. Figure~\ref{fig:input_length} illustrates the variation of the Recall@5 metric under different input sequence lengths.

Experimental results indicate that as the input sequence length increases, the model’s performance in terms of NDCG@5 gradually improves and reaches its peak at $n = 100$. This observation suggests that moderately extending the historical check-in sequence helps the model better capture users’ periodic behavioral patterns.

However, when the trajectory length exceeds 100, a noticeable decline in NDCG@5 is observed. This performance degradation can be attributed to the following factors:

\begin{itemize}
    \item \textbf{Noise interference:} Excessively long input sequences are more likely to include noisy or irrelevant historical information, which can obscure informative patterns and hinder effective feature extraction.
    \item \textbf{Model complexity and overfitting:} Increasing the sequence length substantially raises computational complexity and the risk of overfitting, thereby negatively affecting the model’s generalization performance.
\end{itemize}

\section{Conclusion}

This paper proposes a dual-stream spatio-temporal attention model based on semantic embeddings for next-location prediction. By performing semantic abstraction of trajectory data, functionally decoupling long-term and short-term behavioral patterns, and introducing a context-aware dynamic fusion mechanism for external environmental features, the proposed model achieves a fine-grained characterization of complex user decision-making processes and significantly improves predictive performance.

Comprehensive experiments conducted on large-scale real-world datasets demonstrate that the proposed model consistently outperforms a wide range of baseline methods across key evaluation metrics. Moreover, the dual-stream architecture allows flexible deployment under different business requirements, enabling collaborative handling of scenarios that demand both high timeliness and high-precision prediction. As such, the proposed approach provides a practical and effective technical solution for the refined operation of intelligent mobility platforms.

Despite explicitly considering personalized preferences during the fusion stage, the current model still relies on fixed time-window hyperparameters to partition long-term and short-term sequences. This design may not fully adapt to the diverse habit-formation cycles of different users. In future work, we plan to explore adaptive mechanisms, such as gating strategies or learnable attention modules, to dynamically generate soft partition masks for individual users, thereby enabling more flexible and personalized modeling of user intent evolution.

\bibliographystyle{named}
\bibliography{references}

\end{document}